%Paper: astro-ph/9406059
%From: Andrew P. Gould <gould@payne.mps.ohio-state.edu>
%Date: Tue, 21 Jun 94 17:23:26 EDT

\input phyzzx
\hoffset=0.375in
\overfullrule=0pt
\twelvepoint
\font\bigfont=cmr17
\def\l{\lambda}
\def\p{\partial}
\def\gaml{\Lambda}
\def\max{{\rm max}}
\centerline{\bigfont Analytic Error Estimates}
\bigskip
\centerline{\bf Andrew Gould}
\smallskip
\centerline{Dept of Astronomy, Ohio State University, Columbus, OH 43210}
\smallskip
\centerline{E-mail gould@payne.mps.ohio-state.edu}
\bigskip
\centerline{\bf Abstract}
\singlespace
%\doublespace
	I present an analytic method for estimating the errors in
fitting a distribution.	 A well-known theorem from statistics gives
the minimum variance bound (MVB) for the uncertainty in estimating a
set of parameters $\l_i$, when a distribution function $F(z;\l_1 ... \l_m)$
is fit to $N$ observations of the quantity(ies) $z$.  For example, a
power-law distribution (of two parameters $A$ and $\gaml$) is
$F(z;A,\gaml) = A z^{-\gaml}$.  I present the MVB in a form which is
suitable for estimating uncertainties in problems of astrophysical interest.
For many distributions, such as a power-law distribution or an exponential
distribution in the presence of a constant background, the MVB can be
evaluated in closed form.  I give analytic estimates for the variances
in several astrophysical problems including the gallium solar-neutrino
experiments and the measurement of the polarization induced by a weak
gravitational lens.  I show that it is possible to make significant
improvements in the accuracy of these experiments by making simple
adjustments in how they are carried out or analyzed.   The actual
variance may be above the MVB because of the form of the distribution
function and/or the number of observations.  I present simple methods
for recognizing when this occurs and for obtaining a more accurate
estimate of the variance than the MVB when it does.

Subject Headings: gravitational lensing -- methods: statistical  -- solar
neutrinos

\endpage

\chapter{Introduction}
\normalspace

	A very general problem is to fit a set of observations to a
distribution function of one or several parameters.  For example, one
might want to fit the observed projected separation of binaries, $s$,
to a power-law function of the form, $f(s) d s = A s^{-\ell}$.  In this
case there are two parameters, the index $\ell$ and the normalization $A$.
Or one might fit a point spread function to a two-dimensional gaussian
of position $(x,y)$
which has 6 parameters (1 for normalization, 2 for first moments, 3 for second
moments).  For any given data set, one can always estimate the errors
using monte carlo or other techniques.  However, it is often desirable
to estimate the errors in advance of the observations in order to
optimize the observing program.  Moreover, even if the final error
estimate is made by monte carlo, one would often like to have an
external analytic check on the estimate.

	A well-known theorem in statistics (e.g.\ Kendall, Stuart, \&
Ord 1991) gives a minimum variance bound (MVB) for any linear combination
of the parameters.  It is often the case that the actual variance
is equal to the MVB and that the MVB can be evaluated in closed form.
This makes the MVB potentially a very useful tool.  Nevertheless, standard
texts do not generally make clear how to evaluate this bound in
practical problems nor do they give guidance on how to determine if
the actual variance is roughly equal to the MVB nor how to estimate
the variance when it is well above the MVB.

	Here I present the MVB in a form that allows its direct evaluation.
I present several applications of astrophysical interest for which the
MVB can be evaluated in closed form.  I give a derivation of the MVB
which allows one to see explicitly the conditions under which the
true variance exceeds the minimum.  When this happens, it is usually
possible to modify the MVB formula to give a reasonable estimate of the
true variance.

Suppose that one makes $N$ observations from a distribution $F$ which
is a function of $m$ parameters $\l_1 ... \l_m$ and $p$
known quantities which latter I collectively label $z$.  In the first
example given above $p=1$ and $z=s$.  In the second,
$p=2$ and $z= (x,y)$.  Then in the large
$N$ limit and subject to certain conditions, the covariance between
the  measurements of $\l_i$ and $\l_j$ is given by
$${\rm cov}(\l_i,\l_j) = c_{ij},\qquad c\equiv b^{-1}.\eqn\cdef$$
where
$$b_{ij}\equiv N\VEV{{\p \ln F\over \p \l_i}\ {\p \ln F\over \p \l_j}},
\eqn\bdef$$
and where
$$\VEV{g} \equiv \int d^p z\, F(z;\l_1 ... \l_m) g(z)\bigg/
\int d^p z\, F(z;\l_1 ... \l_m).\eqn\vevdef$$
The variances of the $\l_i$ are then just the diagonal elements of $c$,
${\rm var}(\l_i) = c_{ii}$.

	In \S\ 3, I prove this formula in the limit of large $N$.
In \S\ 4, I show how to modify the formula when it diverges.  I also
show that if $N$ is not large, then
one must smooth $F$ over
the sampling scale before differentiating or alternatively, exclude from
the integrals those parts of observation space that are not well sampled.
First, I present a few  applications.

\chapter{Examples}

\section{Functions With Normalization Plus One Parameter}

	 Some of the most important applications of equation \cdef\
are for distributions
of one variable which have a normalization plus one other parameter.  That is,
$$ F(z;\l_1,\l_2) = A f(z;\gaml),\eqn\simpcase$$
with $\l_1\equiv A$ and $\l_2\equiv\gaml$.  Then
$$b_{11} = {N\over A^2},\qquad b_{12}=b_{21} = {N\over A}
\VEV{d\ln f \over d \gaml}, \qquad
b_{22} = N\VEV{\biggl({d\ln f \over d \gaml}\biggr)^2}.
\eqn\simpbeval$$
The variance in the parameter $\gaml$ is then
$${\rm var}(\gaml) = {1\over N}
\Biggl[\VEV{\biggl({d\ln f \over d \gaml}\biggr)^2}
- \VEV{d\ln f \over d \gaml}^2\Biggr]^{-1}.\eqn\simpvar$$

\section{Power Law}

Suppose that $F$ is a power law, $F(z) = A f(z)$ with $f(z)\equiv z^{-\gaml}$.
And suppose that the observations cover a range $z_1<z<z_2$.  Then
$d\ln f/d z = - \ln z$.  It is straight forward to substitute this expression
into equation \simpvar\ and to evaluate the resulting integrals:
$$\eqalign{{\rm var}(\gaml) = & {1\over N}\biggl[{1\over (\gaml-1)^2}
-{r^{\gaml-1}(\ln r)^2\over (r^{\gaml-1} - 1)^2}\biggr]^{-1}\qquad (\gaml
\not=1)\cr
= & {12\over N}\,(\ln r)^{-2},\qquad(\gaml=1),}\eqn\poweval$$
where $r\equiv z_2/z_1$ and the evaluation is to be done at the best-fit
value of $\gaml$.

\section{Exponential Law In A Cone}

Suppose that one wants to fit the distribution of stars observed in a
cone (say toward the north galactic pole) to an exponential.  (I have
previously solved this problem by more cumbersome methods,
Gould 1989). Taking
account of the volume element as a function of distance, one finds
$f(z) = z^2\exp(-\gaml z)$.  Then $d \ln f/d \gaml = -z$.
Suppose that the observations cover a range from 0 to $z_*$.
Then
$${\rm var}(\gaml) = {1\over N}\biggl(\VEV{z^2}-\VEV{z}^2\biggr)^{-1}
= {\gaml^2\over N}\biggl[12{Q_4(\gaml z_*)\over Q_2(\gaml z_*)}
-9\biggl[{Q_3(\gaml z_*)\over Q_2(\gaml z_*)}\biggr]^2\biggr]^{-1}
,\eqn\varexp$$
where
$$Q_r(x) \equiv 1 - \exp(-x)\sum_{i=0}^r {x^i\over i!}.\eqn\qval$$

\section{The Gallium Solar Neutrino Experiments}

	Consider first the general problem of finding the normalization
$A$ to a known function $f(t)$ in the presence of a constant but
unknown background $B$, over an interval of observations $(0,T)$.  That
is, $F(t) = B + A f(t)$.  Substituting into equation \cdef, one finds
$${{\rm var}(A)\over A^2} = \biggl[\int_0^T dt F(t)
-{T^2\over \int_0^T d t/ F(t)}\biggr]^{-1}.\eqn\galone$$
For the special case of $f(t)=\exp(-t/\tau)$, equation \galone\ becomes,
$$ {{\rm var}(A)\over A^2} = \biggl\{A\tau\biggl[1 - {B\over A}\,
{\ln(1 + A/B)\over W}\biggr]\biggr\}^{-1}\qquad
W\equiv 1 - {\tau\over T}\ln(1 + A/B),\eqn\galtwo$$
where I have assumed $\exp(-T/\tau)\ll B/A$.  Note that the quantity $A\tau$
is just the total `signal', so that the expression in square brackets
is the suppression of statistical significance due to the presence of
background.  Even if the background is known {\it a priori}, there
is still a suppression factor given by setting $W\rightarrow 1$.
Also note that in the limit of high background,
${\rm var}(A)/A^2\rightarrow \tau A^2/2 B$.

	Equation \galtwo\ can be applied
to radio-chemical solar-neutrino detection experiments, such as
the gallium experiments now being conducted by GALLEX
(Anselmann et al.\ 1993; Anselmann et al.\ 1994) and by SAGE
(Abazov et al.\ 1991).  In this case
the total number of expected events, $A\tau$, is a function of the length
of time, $\Delta t$, that the gallium is left in the detector before
being extracted.  That is,
$$A = Q[1-\exp(\Delta t/\tau)].\eqn\qdef$$
where $Q$
is a constant that depends only on the (unknown) flux of solar neutrinos and
the (known) characteristics of the apparatus.  Equation \galtwo\ can
be used to determine the optimal exposure time $\Delta t$ and counting
time $T$ which maximize the efficiency of the experiment.  For illustrative
purposes, I will assume that $T$ is fixed
at $T=180\,$days.  The total length of the experiment, $P$ is fixed.
The total number of runs is then $P/(\Delta t + t_*)$, where $t_*$
is the time required to do the extraction before a new run can be started.
If $\Delta t$ is chosen to be small, each run will constrain the unknown
rate $Q$ relatively weakly, but there will be many runs.  On the other hand,
if $\Delta t$ is large, there will be better constraints from each run, but
fewer runs.  The statistical power of the experiment (inverse square
fractional error in $Q$) is given by
$${\tilde N} = {Q^2\over {\rm var}(Q)} = {P\over \Delta t + t_*}\,
  A\tau\biggl[1 - {B\over A}\,{\ln(1 + A/B)\over W}\biggr].\eqn\galthree$$
In practice, the signal is measured by two sets of counters, one
sensitive to capture of K electrons and the other to capture of L electrons.
The two counters each have their own backgrounds $(B_K$ and $B_L$) which
are different for each run.  They also have their own efficiencies which
lead to slightly different values of $Q_K$ and $Q_L$.  The total statistical
power of the experiment is found by adding the statistical powers from
each of these channels.  To make my estimates for the GALLEX experiment,
I adopt parameters (Anselmann et al.\ 1994; P.\ Anselmann 1994, private
communication)
$\tau=16.5\,$days, $t_*=1.5\,$days, $P=4\,$yr.
$B_K=0.02\,\rm day^{-1}$,
$B_L=0.06\,\rm day^{-1}$, $Q_K=0.11\,\rm day^{-1}$, and
$Q_L=0.12\,\rm day^{-1}$.
I find that the peak value at $\Delta t=14\,$days is $\tilde N=80$ which is
$\sim 13\%$ higher
than the value at $\Delta t=27\,$days, the exposure time currently adopted
by the GALLEX experiment.  I find the optimal exposure
time for the SAGE experiment is also about 2 weeks.

\section{Centroid of a Distribution}

	Suppose that a distribution is a known, circularly symmetric
function around an unknown
coordinate center $(X,Y)$.  That is $F(x,y;X,Y) = A f(r)$, with
$r^2\equiv (x-X)^2 +(y-Y)^2$.  In this case, there are three parameters,
$\l_1=A$, $\l_2=X$, and $\l_3=Y$.  The problem greatly simplifies, however,
because all the off-diagonal elements of $b_{ij}$ vanish.  Thus,
${\rm var}(X)= 1/b_{22}$, or
$${\rm var}(X) = {\rm var}(Y) = {2\over N}\VEV{\biggl({d\ln f\over d r}
\biggr)^2}^{-1}.\eqn\varxy$$

	From equation \varxy, one immediately finds that for a gaussian
${\rm var}(X) = \sigma^2/N$, where $\sigma$ is the standard deviation
of the distribution.  One might want to generalize from this to infer that the
error in the centroid always scales as the standard deviation of the
distribution.  However, for distributions
more sharply peaked than a gaussian
the variance can be much smaller than this naive estimate.

\section{Statistics of Weak Gravitational Lenses}

	An important example of centroiding a sharply peaked distribution
arises in the statistics of weak gravitational lensing of distant galaxies.
The observed ellipticity of each galaxy can be represented by a 2-dimensional
vector whose magnitude is the ratio of the difference to the sum of the major
and minor axes and whose direction is twice the position angle of the major
axis.  The effect of a weak gravitational lens is to change the ellipticity
of an image ${\bf e}_i$ relative to that of the object ${\bf e}_o$
by a 2-dimensional vector $\bf p$, called the polarization.
That is, ${\bf p = e}_i - {\bf e}_o$.
The mean polarization in a given region
of the sky can be determined by comparing the centroid of the
ellipticities of the galaxy images with the centroid of the object
ellipticities, which latter is assumed to be statistically consistent with
zero.

	I take the underlying true distribution of ellipticities, $\bf e$, to
be
$$g({\bf e}) = {e_\max - e\over \pi e_\max^2 e},\qquad 0<e<e_\max=0.8
,\eqn\emax$$
consistent with the ellipticities observed by Mould et al.\ (1994) after
convolution with their measurement errors (J.\ Villumsen 1994, private
communication).  I take the observed distribution (in the absence of lensing),
$f(e)$, to be this underlying distribution convolved with a gaussian
measurement error, $\sigma$.  I then use equation \varxy\ to
evaluate the error $\delta p$ in determining each component of the
polarization.
\FIG\two{Statistical power (inverse square error per galaxy observed) for
measurement of the polarization due to weak gravitational lensing ({\it solid
line}). The numerical evaluation is made by convolving eq.\ \emax\ with
a gaussian error with standard deviation $\sigma$ and substituting the result
into eq.\ \varxy.  Also shown ({\it dashed line}) is the statistical power if
one uses equal weighting, the customary method of finding the centroid of
the ellipticities of an observed set of galaxies.}
The statistical power (inverse variance per galaxy observed)
as a function of $\sigma$ is shown by the solid line in Figure \two.
Also shown
(dashed line) is the statistical power
if the polarization is estimated by taking
the unweighted sum of the galaxy ellipticities, a procedure which is
customary in virtually all observational studies (e.g.\ Fahlman et al.\ 1994;
Mould et al.\ 1994; Smail et al.\ 1994) and most theoretical studies
as well.  Note that if the errors are fairly large $(\sigma\gsim 15\%)$,
then the naive approach of equal weighting is almost as good as the MVB
obtained if one uses maximum likelihood.  The reason for this is that
if $\sigma$ is large, the central portion  of the distribution $f(e)$ is
dominated by gaussian errors and so is nearly gaussian.  It is straight
forward to show that when fitting the centroid of a gaussian distribution,
maximum likelihood is equivalent to equally weighting the data.
Note also
however, that if the errors are relatively small $(\sigma\lsim 5\%)$, then
equal weighting wastes a large fraction of the information available
in the data.

\chapter{Derivation}

	Here I prove equation \cdef.
	Consider a small patch of observation space of volume
$(\Delta z)^p$ around a point $z$.  Suppose that the parameters
$\l_1^* ... \l_m^*$ have been chosen so that they are at or very near the
best fit.  The predicted number of observations falling within the volume is
$n_{\rm pred} = F(z;\l_1^* ... \l_m^*)(\Delta z)^p$, while the number
actually observed is a possibly different number, $n_{\rm obs}$.  I work in the
limit of large $n$, so that $|(n_{\rm pred}/n_{\rm obs})-1|\ll 1$.
I form $\chi^2$ by summing over all such patches
$$\chi^2 = \sum_k { [(\Delta z)^p F(z_k;\l_1^* ... \l_m^*)
-n_{{\rm obs},k}]^2\over
n_{\rm pred},k}.\eqn\chithr$$
I then linearize the equation in the neighborhood of the $\l_j^*$ to obtain
a final correction $\Delta \l_j$,
$$\eqalign{\chi^2 = &\sum_k {[(\Delta z)^p
\sum_i \Delta \l_i \p F(z_k;\l_1^* ... \l_m^*)/\p \l_i
- y_k]^2\over
n_{{\rm pred},k}},\cr y_k\equiv & n_{{\rm obs},k}-F(z_k;\l_1^* ... \l_m^*)
(\Delta z)^p.}\eqn\chione$$
To minimize $\chi^2$, I differentiate and set $d\chi^2/d\Delta \l_i = 0$ which
yields a set of linear equations for the final correction $\Delta \l_j$
to the initial estimates $\l_j^*$,
$$ d_i = \sum_j b_{ij} \Delta \l_j$$
where
$$\eqalign{d_i = &
\sum_k {(\Delta z)^p\over n_{{\rm pred},k}}
{\p F(z_k;\l_1 ... \l_m)\over \partial l_i}y_k,\cr
\qquad b_{ij} = &\sum_k {(\Delta z)^{2p}\over n_{{\rm pred},k}}
\biggl({\p F(z_k;\l_1 ... \l_m)\over \p \l_i}\biggr)\,
\biggl({\p F(z_k;\l_1 ... \l_m)\over \p \l_j}\biggr).}\eqn\chitwo$$
The solution is given by $\Delta \l_i= \sum_j c_{ij} d_j$, where
$c\equiv b^{-1}$.  More important from the present perspective,
the covariance of $\Delta \l_i$ with $\Delta \l_j$
(and hence the covariance of $\l_i$ with $\l_j$) is given by $c_{ij}$.
(See Press et al.\ 1986).  I evaluate $b_{ij}$ by
converting the sum to an integral and find
$$b_{ij} = \int d^p z {1\over F(z;\l_1 ... \l_m)}
{\p F\over \p\l_i} {\p F\over \p\l_j}
= N\VEV{ {\p \ln F\over \p\l_i} {\p \ln F\over \p\l_j}}.\eqn\prf $$

\chapter{Range of Validity}

	There are some distribution functions for which equation \cdef\ is
obviously not valid.  Consider, for example the distribution
$$F(z) = A(\gaml -z),\eqn\pathone$$
over the interval $(0,\gaml)$.  Using the formalism
of \S 2, I find
${\rm var}(\gaml) =\{N [\VEV{(\gaml- z)^{-2}} - \VEV{(\gaml -z)^{-1}}^2
]\}^{-1}$.
The second term is finite, but the first is logarithmically divergent.
The problem is that in the proof, I assumed that there are a large number
of observations in every patch of observation space.  However, in this
example the expected number of observations in the interval $(\gaml-
\Delta z,\gaml)$ is $N(\Delta z)^2/\gaml$.  For
$\Delta z< (\gaml/N)^{1/2}$,
this number is smaller than unity, so that the assumptions underlying the
proof break down.  In essence, the formal derivation treats this region
of parameter space as contributing an infinite amount of information
whereas it actually contributes almost none.  A practical
solution is to cut off the integral at $z=\gaml-(\gaml/N)^{1/2}$,
which results in
the estimate ${\rm var}(\gaml) = (\gaml^2/N)/\ln(N\gaml e^{-4})$.
The choice of the cutoff is somewhat arbitrary,
but the error in the estimate is logarithmic in the cut-off value.

	Even for distributions which give rise to finite
integrals one may still ask how large the number of observations must
be to reach the `limit of large $N$'.  For small $N$, one usually fits
a distribution by maximum likelihood (ML) rather than the binned $\chi^2$
method that I modeled in the previous section.  (Indeed, it is common
to use ML even with large $N$.  However, as I show below, the two
methods are equivalent for large $N$.)\ \ Thus, to address this question,
I begin by examining the relation between ML and binned $\chi^2$ fits.

	In the large $N$ limit, the probability of observing $n_{{\rm obs},k}$
given $n_{{\rm pred},k}$, is gaussian distributed.  Hence, up to an
irrelevant constant, $-\chi^2/2$ is the logarithm of the probability of
making the observations given the model.  One could imagine further
sub-dividing each bin of volume $(\Delta z)^p$ into ${\cal N}$ equal sub-bins,
with ${\cal N}\gg n_{{\rm pred},k}$.  The predicted number in each bin,
 $\tau_k= n_{{\rm pred},k}/{\cal N}\ll 1$, would then be far in the Poisson
limit, and would be distributed as
$$P(n;\tau) = {\tau^n\over n!}\exp(-\tau).\eqn\pois$$
Nevertheless, the joint probability of observing one event in each of
$n_{{\rm obs},k}$ sub-bins and observing nothing in the remaining
${\cal N}-n_{{\rm obs},k}$ sub-bins is identical to the probability of
observing $n_{{\rm obs},k}$ in the whole bin and in particular is
gaussian distributed.  The reason that the extreme Poisson
(and hence non-gaussian) probabilities combine to form a gaussian joint
distribution is that {\it the individual probability functions in the sub-bins
are identical}.  This identity arises from the fact that the distribution
function is effectively constant over the bin.  Hence, in the large $N$
limit (and up to an irrelevant constant) one can rewrite
${\chi^2/ 2} = - \ln L + {\rm const}$, where $L$ is a product over
extremely small bins of volume $(\delta z)^p$,
$$L = \prod_k P(n'_{{\rm obs},k};\tau_k),
\qquad (n'_{{\rm obs},k}=0\ {\rm or }\ 1),
\eqn\Ldef$$
and where
$$\tau_k = (\delta z)^p F(z_k;\l_1 ... \l_m).\eqn\taudef$$
In the extreme Poisson limit $n!=1$, so that $\ln P(n;\tau)= n\ln\tau - \tau$.
Hence,
$$\eqalign{\ln L = &\sum_k n'_{{\rm obs},k}\ln [F(z_k)(\delta z)^p]
- \sum_k F(z_k)(\delta z)^p\cr = &\sum_{{\rm obs},k}\ln F(z_k)
+ N_{\rm obs}\ln(\delta z)^p - N_{\rm pred}.}\eqn\Lone$$
where $N_{\rm pred}$ and $N_{\rm obs}$ are respectively the predicted and
actual number of observations.  The second term depends only on the bin
size and not the model and so is irrelevant.  If one restricts $F$ to
a class of functions with the same normalization, then the last term
is also irrelevant.  One then arrives at the standard likelihood formulation
as a sum of the log probabilities over the observed events,
$$\ln L = \sum_{{\rm obs},k} \ln F(z_k;\l_1 ... l_m).\eqn\Ltwo$$
In the large $N$ limit, maximizing this likelihood function is equivalent
to minimizing $\chi^2$.  (It also has the convenient advantage that no binning
is required and for that reason is often used instead of $\chi^2$.)\ \
The proof given in the
previous section therefore also applies to distributions fit by ML in the
large $N$ limit.  When $N$ is not large or if $F$ is pathological, the
proof breaks down and some care is required.

Equation \pathone\ is clearly one example of a pathological
distribution, since the integral $\VEV{(d\ln f/d\gaml)^2}$ diverges.
What makes the integral pathological is that it has a very large
(in this case, infinite) contribution from a region of
observation space that in practice contributes very little information
to the estimate.  The integral was derived by assuming gaussian statistics
over this region.  From the above derivation of the ML formula, it is clear
that this assumption is valid only in regions where there are a large
number of observations with identical (or in practice, similar) probability
distributions.  If equation \pathone\ were convolved with a gaussian
of very small width, say $\gaml/100$, then  $\VEV{(d\ln f/d\gaml)^2}$ would
be formally convergent.  However, equation \bdef\ would still underestimate
the errors unless $N$ were large enough to probe the regions making
large contributions to the integral.  In general, then, equation \bdef\
will yield the correct errors if $F$ is first smoothed on the scale
of the sampling.  If $F$ is already smooth on these scales, then equation
\bdef\ will be essentially correct.

	To get a practical sense of these requirements, consider the
problem of the determining the weak lensing polarization as discussed
in \S\ 2.6.  Suppose that the error in the ellipticity measurement were
$\sigma=2\%$.  If the ellipticities of $N=12$ galaxies were measured then
according to Figure \two, the error in the polarization would be
$\delta p\sim 1/30$.  However, there would be only $\sim 1$ galaxy
with a measured polarization within $1/30$ of the centroid.  Hence, it
is implausible that the centroid could be measured this accurately.
Unfortunately, in this case the integral \varxy\ is power-law divergent
near 0 (until it is cut off by the observational errors) so the estimate
of the variance is sensitive to the choice of a cut off.  If one
were interested in a very accurate estimate of the variance, a monte
carlo simulation would be required.  On the other hand, suppose that
$N=300$.  In this case Figure \two\ predicts $\delta p\sim 1/150$.
The distribution function is smooth on scales of $\sim 2\%$ and there
are ${\cal O}(10)$ galaxies within 0.02 of the centroid.  Thus, the MVB
would provide a reasonable estimate.

{\bf Acknowledgements: }  I would like to thank R.\ Lupton for making
several helpful suggestions.

\endpage
\Ref\ab{Abazov, A.\ I.\ 1991, Phys.\ Rev.\ Lett., 67, 3332}
\Ref\ansone{Anselmann, P., et al.\ 1993, Phys.\ Lett.\ B, 314, 445}
\Ref\anstwo{Anselmann, P., et al.\ 1993, Phys.\ Lett.\ B, submitted}
\Ref\gould{Gould, A.\ 1989, ApJ, 341, 748}
\Ref\fk{Fahlman, G.\ G., Kaiser, N., Squires, G., \& Woods, D.\ 1994,
ApJ, submitted}
\Ref\kend{Kendall, M., Stuart, A.\ \& Ord, T.\ K.\ 1991,
Advanced Theory of Statistics, p.\ 623 (New York: Oxford Univ.\ Press)}
\Ref\press{Press, W.\ H., Flannery, B.\ P., Teukolsky, S.\ A., \&
Vetterling, W.\ T.\ 1986, Numerical Recipes (Cambridge: Cambridge Univ.\
Press)}
\Ref\aa{Mould, J., Blandford, R., Villumsen, J., Brainerd, T.,
Smail, I., Small, T., \& Kells, W.\ 1994, MNRAS, submitted}
\Ref\smaila{Smail, I., Ellis, R.\ S., Fitchett, M.\ J., \& Edge, A.\ C.\ 1994,
MNRAS, submitted}
\refout
\endpage
\figout
\endpage
\bye